\documentclass[]{interact}
\usepackage{graphicx} 
\usepackage{amsmath}
\usepackage{svg}
\usepackage{algorithm}
\usepackage{algpseudocode}
\usepackage{nicefrac}

\usepackage{tikz}

\usepackage{epstopdf}
\usepackage{subfigure}

\usepackage{natbib}
\bibpunct[, ]{(}{)}{;}{a}{}{,}
\renewcommand\bibfont{\fontsize{10}{12}\selectfont}

\theoremstyle{plain}
\newtheorem{theorem}{Theorem}[section]

\theoremstyle{definition}

\newtheorem{example}[theorem]{Example}

\theoremstyle{remark}

\begin{document}

\articletype{ARTICLE TEMPLATE}
\title{The formula for the completion time of project networks}


\author{
\name{Manuel Castejón-Limas\thanks{CONTACT M.~Castejón-Limas.
Email: mcasl@unileon.es}, Gabriel Medina Martínez, Virginia Riego del Castillo and Laura Fernández-Robles}
\affil{Departamento de Ingenierías Mecánica, Informática y Aeroespacial, Universidad de León, Spain}
}

\maketitle

\begin{abstract}
This paper formulates the completion time $\tau$ of a project network as $ \tau =\|\mathbf{R} \mathbf{t} \|_\infty $ where the rows of $\mathbf{R}$ are simple paths of the network and $\mathbf{t}$ is a column vector representing the duration of the activities. Considering this product as a linear transformation leads to interesting findings on the topological relevance of both paths and activities using singular value decomposition. The notion of spectral networks is introduced to condense the fundamental structure of the project network. A definition of project stress is introduced to establish a comparison index between two alternatives in terms of slack. Additionally, the Moore-Penrose inverse of $\mathbf{R}$ is presented to find the configuration of the durations of the activities resulting in a given simple path duration vector.  Then, the systematic mapping review process carried out to assess our claims' novelty is reported. Finally, we reflect on the notion of relevance for paths and activities and the relationship of the incidence matrix with the proposed approach.
\end{abstract}

 
\section{Introduction}
This paper adopts the representation of a project network as a graph with activities on arrows. What follows can be equally applied to the activities on node representation.
\subsection{Glossary of symbols}
We use the following symbols:
\begin{description}
    \item[$a_{ij}$:] Coefficient that indicates if node $n_i$ is the source of activity $A_j$ (positive 1), the sink of $A_j$ (negative 1), or else (0 value) $A_j$ is not related to $n_i$.

    \item[$A_i$:] The $i$-th activity
    
    \item[$b_i$:] Coefficient that indicates if node $n_i$ is an intermediate node (0 value), the project's start node (1 value), or the project's finishing node (-1 value).

    \item[$\mathbf{b}$:] Column vector collecting $b_i$ coefficients

    \item[$c_i$] $i$-th coefficient in a linear combination

    \item[$\boldsymbol{\delta}$:] Vector in the nullspace of matrix $\mathbf{R}$

    \item[$e_i$:] Early time of node $n_i$.

    \item[$\mathbf{G_i}$:] Spectral matrix $\mathbf{G_i} = \lambda_i \mathbf{u_i} \mathbf{v_i}$ 

     \item[$\mathbf{H}$:] Incidence matrix of a project network

    \item[$\lambda_i$:] $i$-th singular value
    
   \item[$n_i$:]  $i$-th node

   \item[$N\left( \mathbf{R} \right) $:] Nullspace of matrix $\mathbf{R}$

   \item[$\mathbf{R}$:] Route matrix

    \item[$\mathbf{R}^+$:] Moore-Penrose inverse of $\mathbf{R}$

   \item[$\mathbf{\Sigma}$:] Diagonal matrix containing the singular values
   
   \item[$S_p$] Project stress using norm $L_p$
   
   \item[$\tau$:] Completion time of the project

   \item[$\tau_i$:] Completion time of the $i$-th simple path
   
   \item[$\boldsymbol{\tau}$:] Column vector collecting the $\tau_i$
   
   \item[$t_i$:] Duration of activity $A_i$
   
   \item[$\mathbf{t}$:] Column vector containing the duration of the activities

   \item[$\mathbf{u_i}$:] $i$-th column vector of $\mathbf{U}$

   \item[$\mathbf{U}$:] Matrix of left singular vectors

   \item[$\mathbf{v_i}$:] $i$-th row vector of $\mathbf{V}^T$

   \item[$\mathbf{V}$:] Matrix of right singular vectors

   \item[$x_i$:] Binary variable indicating whether $A_i$ belongs to the critical path
   \end{description}
   We define a simple path as the linear sequence of activities that begins at the project's starting node and ends at its finishing node. Lastly, we use SVD as the acronym for Singular Value Decomposition.

\subsection{Project completion time using linear programming optimization}
The problem of estimating the completion time of a project network can be posed as a linear programming optimization problem \citep{Klerides2010, Chen2008}:
\[
\text{Maximize } \tau = t_1x_1 + t_2x_2 + \cdots + t_nx_n
\]
\textbf{Subject to:}
\[
\begin{aligned}
    a_{11}x_1 &+ a_{12}x_2 + \cdots + a_{1n}x_n &= b_1 \\
    a_{21}x_1 &+ a_{22}x_2 + \cdots + a_{2n}x_n &= b_2 \\
    &\vdots \\
    a_{m1}x_1 &+ a_{m2}x_2 + \cdots + a_{mn}x_n &= b_m \\
    x_1, & \, x_2,\, \ldots,\, x_n\, & \geq 0
\end{aligned}
\]
where $t_j$ stands for the duration of activity $A_j$, $x_j$ are the basic variables that allow the optimization algorithm to choose amongst different potential paths, and the coefficients $a_{ij}$ and $b_{i}$ are either $0$, $1$ or $-1$ to represent the concrete structure of the project network and comply with conservation laws.

This linear programming problem for optimizing the completion time can be solved using iterative algorithms such as simplex.

\begin{example}
Consider the project example shown in Figure \ref{fig:Ejpert}. We will repeatedly refer to this minimal example as `the toy project'.

\begin{figure}[h]
    \centering
    \includegraphics[width=\textwidth]{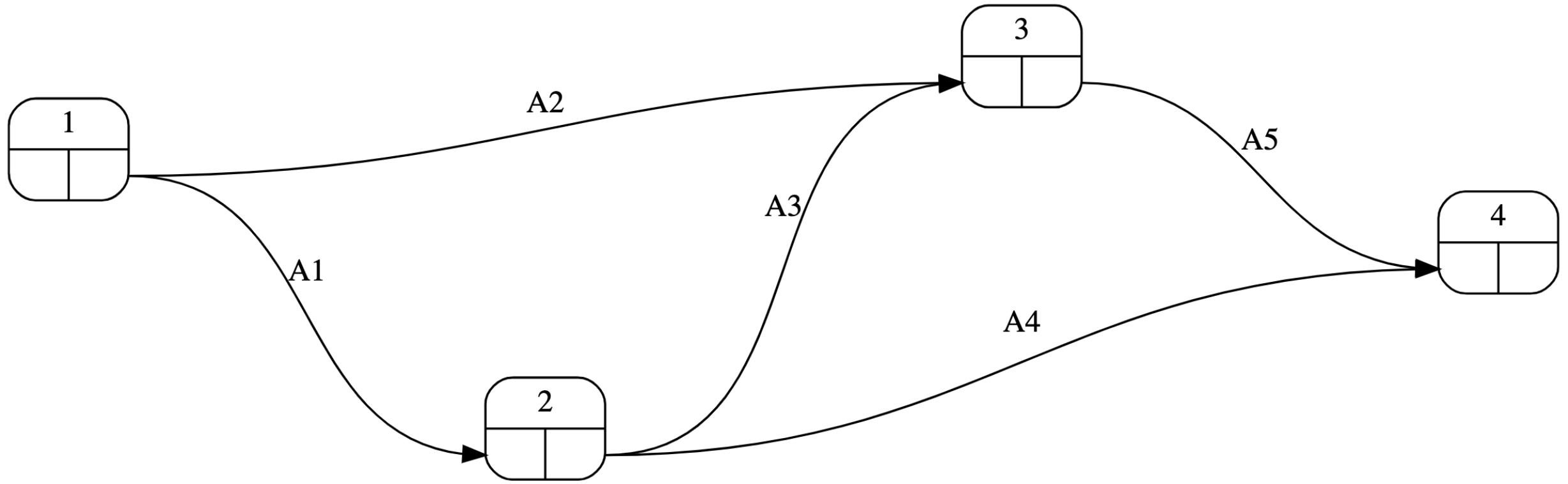}
    \caption{Illustrative toy example of a project network}
    \label{fig:Ejpert}
\end{figure}

The linear programming optimization for the toy project can be defined as:
\[
\text{Maximize } \tau = t_1x_1 + t_2x_2 + t_3x_3 + t_4x_4 + t_5x_5
\]
Subject to:
\[
\begin{aligned}
    x_1 &+ x_2 &= 1 \\
    -x_1 &+ x_3 + x_4 & = 0 \\
    -x_2 &- x_3 + x_5 & =  0 \\
    -x_4 &- x_5 & =  1 \\
    x_1, & \, x_2,\, x_3,\, x_4, \, x_5\, & \geq 0
\end{aligned}
\]
\end{example}
\subsection{Project completion time using a heuristic approach}
The optimization problem is nevertheless rarely used as the completion time can be obtained by the traditional alternative heuristic algorithm \citep{Qi2012}:
\begin{algorithmic}[1]
\State Assign 0 to the early time of the start node.
\State Visit all nodes in topological order calculating the early time $e_j$ of node $n_j$ as $e_j=\max\{e_i + t_{ij}\}$ for all activities $A_{ij}$ starting on node $n_i$ and finishing on node $n_j$.
\State The early time of the finishing node is the project's completion time.
\end{algorithmic}

\begin{example}
For the toy project, the iterative approach for calculating the early times of the nodes, and thus the completion time of the project follows these steps:
\begin{enumerate}
    \item $e_1=0$
    \item $e_2= e_1 + t_1$
    \item $e_3= \max\{ e_1 + t_2 ; e_2 + t_3\}$
    \item $e_4= \max\{ e_3 + t_5 ; e_2 + t_4\}$
\end{enumerate}
\end{example}

\subsection{Project completion time using simple paths}
The topological order followed in the calculation ensures that the values of all the variables required for the computation of the early time of a particular node have been previously computed and the calculation can proceed numerically without dragging symbolic references throughout the forward pass calculation. Nevertheless, should we wish to keep the symbolic approach, we would need to rely on the following properties that the $\max$ function satisfies:
\[
\begin{aligned}
\max\{ m + n \,; p + q \} + r & =  \max\{ m + n + r\,; p + q + r\} \\
\max\{ \max\{ m\,; n\} \,; p \} & =  \max\{ m \,; n \,; p\} 
\end{aligned}
\]
These properties allow the simplification of the early time formulas leading to:
\begin{equation}
\label{eq:maxTj}
\tau = \max_i \{ T_i \}
\end{equation}
where $T_i$ accounts for the duration of the simple paths from the start to the finishing node~\citep{DODIN2006, CraigWSchmidt}.
\begin{example}
For the toy project, we could write the symbolic expressions as follows:
\begin{enumerate}
    \item $e_1=0$
    \item $e_2= t_1$
    \item $e_3= \max\{ t_2 ; t_1 + t_3\}$
    \item $e_4= \max\{ \max\{ t_2\,; t_1 + t_3\} + t_5\,; t_1 + t_4\} = \max\{ \max\{ t_2 + t_5\,; t_1 + t_3 + t_5\}\,; t_1 + t_4\} = \max\{ t_2 + t_5\,; t_1 + t_3 + t_5\,; t_1 + t_4\}$
\end{enumerate}
This result shows that the early time of the finishing node depends on the activities belonging to three different simple paths:
\begin{enumerate}
    \item $R_1 = n_2 \rightarrow n_5$
    \item $R_2 = n_1 \rightarrow n_3 \rightarrow n_5$
    \item $R_3 = n_1 \rightarrow n_4$
\end{enumerate}
The simple path with the longest duration will determine the project completion time.
\end{example}
The rest of the paper is structured as follows: Section \ref{sec:algebraicFormula} defines a formula for the project completion time and elaborates on the insights its algebraic properties can provide. Section \ref{sec:mapping} describes the systematic mapping literature review carried out to assess the novelty of our claims. Section \ref{sec:discussion} reflects on the meaning of the relevance of a path depending on topological, deterministic, and stochastic components. It also relates the linear programming constraint equations with the incidence matrix and discusses the benefits of using the proposed approach instead of one based on the latter. 

\section{The completion time algebraic formula}
\label{sec:algebraicFormula}
Equation \ref{eq:maxTj} compares the values contained in the different $T_j$, and these are linear combinations of the duration of those activities belonging to path $j$. It can be reformulated, in matrix form as:
\begin{equation}
\label{eq:tauIsRt}
\tau =\left|\left|\mathbf{R} \mathbf{t} \right|\right|_\infty
\end{equation}
where $\mathbf{t}$ is a vector containing the duration of the activities, and $\mathbf{R}$ is the route matrix whose cells adopt values 0 or 1 depending on whether the activity $A_i$ in the column participates in the simple path $R_j$ in the row. Alternative names for $\mathbf{R}$ could be `simple path matrix' or just `path matrix'. 
\begin{example}
For the toy project, we have the above-mentioned three simple paths. They are encoded in the rows of the $\mathbf{R}$ matrix as:
\[
\mathbf{R} =  \begin{pmatrix}
0 & 1 & 0 & 0 & 1 \\
1 & 0 & 1 & 0 & 1 \\
1 & 0 & 0 & 1 & 0 \\
\end{pmatrix}
\]
and thus:
\[
\tau =\left|\left| \mathbf{R} \mathbf{t} \right|\right|_\infty = \left|\left| \begin{pmatrix}
0 & 1 & 0 & 0 & 1 \\
1 & 0 & 1 & 0 & 1 \\
1 & 0 & 0 & 1 & 0 \\
\end{pmatrix} \begin{pmatrix} t_1 \\
t_2\\
t_3\\
t_4\\
t_5\\ 
\end{pmatrix}
 \right|\right|_\infty = \max\{ t_2 + t_5\,; t_1 + t_3 + t_5\,; t_1 + t_4\}
\]

\end{example}    
Equation \ref{eq:tauIsRt} combines most compactly how the network topology and the duration of the activities interact to provide the completion time of the project; and they do it in the simple product of two independent matrices. 

In Section \ref{sec:mapping}, we conduct a systematic mapping literature review to determine whether Equation \ref{eq:tauIsRt} has been previously referenced in the literature. The review revealed that only one source, \cite{CraigWSchmidt}, presented a matrix fundamentally similar to our proposal. However, their matrix included extra rows (dummy paths with a single activity) and columns (dummy activities) as needed to convert it into a square matrix for tractability purposes.

Our primary assertion in this paper is that our definition of $\mathbf{R}$ is a significant advancement. This simpler rectangular definition offers valuable insights without the extraneous noise introduced by the additional dummy paths and activities.

\subsection{Nullspace of $\mathbf{R}$}
Having an algebraic formula instead of algorithms allows for elaborating on the implications of Eq. \ref{eq:tauIsRt}.
The most obvious issue is that the product $\mathbf{R} \mathbf{t}$ can be algebraically interpreted as a linear map, and thus involve what linear algebra has to say. We begin by considering the nullspace of matrix $\mathbf{R}$, in what follows $N\left(\mathbf{R}\right)$, the vector space that satisfies Eq. \ref{eq:nullspace}.
\begin{equation}
\label{eq:nullspace}
\mathbf{R} \boldsymbol{\delta} = \mathbf{0}
\end{equation}

The scenario where $\boldsymbol{\delta}$ represents an increment to the duration of the activities is of particular interest. These values of $\boldsymbol{\delta}$ lead to different configurations of the activity durations, $\mathbf{t}_1$ and $\mathbf{t}_2$ in Eq. \ref{eq:nullDelta}, that maintain the overall project completion time unchanged, even though the duration of individual activities may vary.
\begin{equation}
\label{eq:nullDelta}
\mathbf{R} \mathbf{t}_2 =\mathbf{R} \left( \mathbf{t}_1 + \boldsymbol{\delta}  \right) = \mathbf{R}  \mathbf{t}_1 + \mathbf{R} \boldsymbol{\delta} = \mathbf{R}  \mathbf{t}_1 + \mathbf{0} = \mathbf{R}  \mathbf{t}_1
\end{equation}
In the context of optimization analyses, such as those related to resource programming or project crashing, identifying these alternatives provides opportunities to enhance the project's value while keeping the project duration fixed.

\begin{example}
For the toy project, $N\left(\mathbf{R}\right)$ is given by the linear combinations shown in Eq. \ref{eq:NRtoy}.
\begin{equation}
\label{eq:NRtoy}
 \boldsymbol{\delta} = c_1\left[\begin{matrix}-1\\0\\1\\1\\0\end{matrix}\right] + c_2 \  \left[\begin{matrix}0\\-1\\-1\\0\\1\end{matrix}\right]
\end{equation}
where $c_1$ and $c_2$ are arbitrary constants. These values indicate that, for example, increments of $c_1$ in the durations of activities $A_3$ and $A_4$ will result in the same overall project duration, provided that the duration of activity $A_1$ is decremented by the same amount, $c_1$. Similarly, decrements of $c_2$ in the durations of activities $A_2$ and $A_3$ will result in the same overall project duration, provided that the duration of activity $A_5$ is incremented by the same amount, $c_2$. The same result will be obtained by any vector obtained for the linear combination expressed in Eq. \ref{eq:NRtoy}. Given additional information about, say cost, these alternatives might provide a cheaper schedule configuration.
\end{example}

\subsection{Path and activity relevance in the network topology}
The interpretation of Eq. \ref{eq:tauIsRt} as a linear map leads to the second claim of this paper: the singular value decomposition of matrix $\mathbf{R}$ provides valuable information on the topological relevance for project managers of both simple paths and the activities of the project network.
\begin{equation}
\label{eq:svdR}
\mathbf{R} = \mathbf{U} \mathbf{\Sigma} \mathbf{V}^T
\end{equation}
$\mathbf{\Sigma}$ is a diagonal matrix whose ordered elements $\lambda_i$ weigh the impact of the corresponding left and right singular vectors. We will be interested in the largest values as they will be the ones associated with those paths or activities of greater topological relevance. Let's define $\lambda_{max}$ as the largest $\lambda_i$ and $i_{max}$ as its position in the diagonal.
Column $i_{max}$ of $\mathbf{U}$ contains the information on the relevance of each simple path; the row with the largest values in that column indicates the most relevant routes.
Row $i_{max}$ of $\mathbf{V}^T$ contains the information on the relevance of each activity; the column with the largest values in that row indicates the most relevant activities.
\begin{example}
For the toy project, the SVD of $\mathbf{R}$ renders:
\[
\mathbf{R} = \begin{pmatrix}
  -0.4 & 0.7 & 0.6\\
  -0.8 & -0. & -0.6\\
  -0.4 & -0.7 & 0.6\\
\end{pmatrix}  \begin{pmatrix}
  2.0 & & \\
  & 1.4 & \\
  & & 1.0\\
\end{pmatrix} \begin{pmatrix}
 -0.6 & -0.2 & -0.4 & -0.2 & -0.6\\
  -0.5 & 0.5 &  0.0 & -0.5 & 0.5\\
  -0.0 & 0.6 & -0.6 &  0.6 & 0.0\\
\end{pmatrix}
\]
The largest singular value is in position 1. Thus we will consider the first column of $\mathbf{U}$ and the first row of $\mathbf{V}^T$. We find that the largest value in the first column of $\mathbf{U}$ ($-0.8$) is located in the second row, corresponding to route $R_2 = n_1 \rightarrow n_3 \rightarrow n_5$. The largest value in the first row of $\mathbf{V}^T$ ($-0.6$) is located in the first and last columns, corresponding to activities $A_1$ and $A_5$.
Thus, in topological terms, the most relevant route for this particular network is $R_2$ and the most relevant activities are $A_1$ and $A_5$.
\end{example}

\subsection{Spectral theorem applied to project networks}
Now that the project network has been characterized by the SVD of $\mathbf{R}$ we can invoke the spectral theorem to decompose $\mathbf{R}$ as a sequence of network matrices $\mathbf{G_i}$ ordered by their impact $\lambda_i$ on the global network configuration.
\[
\mathbf{R}=\sum_i \mathbf{G_i}=\sum_i \lambda_i\mathbf{u}_i \mathbf{v}_i
\]
where $\mathbf{u}_i$ are the column vectors of $\mathbf{U}$ and $\mathbf{v}_i$ the row vectors of $\mathbf{V}^T$.

We will refer to these $\mathbf{G_i}$ as spectral PERTS, or more generally, spectral networks. The spectral decomposition poses opportunities for understanding the fundamental structure of a project network and thus simplifying its complexity for the sake of better decision-making.
\begin{example}
For the toy project, we have the following spectral networks:
\[
\begin{aligned}
\mathbf{G_1} &=\begin{bmatrix}
  0.5 & 0.2 & 0.3 & 0.2 & 0.5\\
  1 & 0.3 & 0.7 & 0.3 & 1\\
  0.5 & 0.2 & 0.3 & 0.2 & 0.5\\
\end{bmatrix}\\
\mathbf{G_2} &=  \begin{bmatrix}
  -0.5 & 0.5 & 0 & -0.5 & 0.5\\
  0 & 0 & 0 & 0 & 0\\
  0.5 & 0.5 & 0 & 0.5 & -0.5\\
\end{bmatrix}\\
\mathbf{G_3} &= \begin{bmatrix}
  0 & 0.3 & -0.3 & 0.3 & 0\\
  0 & -0.3 & 0.3 & -0.3 & 0\\
  0 & 0.3 & -0.3 & 0.3 & 0\\
\end{bmatrix}\\
\end{aligned}
\]
Filtering those elements less than 0.6 as an arbitrary threshold for this particular network, and rounding, we get:
\[
\begin{aligned}
\mathbf{G_1} &=\begin{bmatrix}
  0 & 0 & 0 & 0 & 0\\
  1 & 0 & 1 & 0 & 1\\
  0 & 0 & 0 & 0 & 0\\
\end{bmatrix}\\
\mathbf{G_1}+\mathbf{G_2} &=  \begin{bmatrix}
  0 & 1 & 0 & 0 & 1\\
  1 & 0 & 1 & 0 & 1\\
  1 & 0 & 0 & 1 & 0\\
\end{bmatrix}\\
\end{aligned}
\]
For the toy project the first spectral network $\mathbf{G_1}$ compresses the structure of the project network into path $R_2$. Enhancing this network with $\mathbf{G_2}$ we get that $\mathbf{G_1}$ + $\mathbf{G_2}$ already represents all details in $\mathbf{R}$.
\end{example}

\subsection{Project stress}
Equation \ref{eq:tauIsRt} expresses the completion time of the project using an $L_\infty$ norm. We can consider the situation where two different duration vectors, $\mathbf{t}_1$ and $\mathbf{t}_2$, provide the same completion time, such that $ \|\mathbf{R} \mathbf{t}_1 \|\infty = \|\mathbf{R} \mathbf{t}_2 \|\infty$. However, under a different norm definition, these two values might differ. The project with the larger norm will have simple paths whose durations are closer to that of the critical path, while the project with the smaller norm will have shorter simple paths.

Following this rationale, we propose a definition for the stress, $S_p$, of a project configuration as shown in Eq. \ref{eq:stress}:

\begin{equation}
\label{eq:stress}
S_p = \frac{ \|\mathbf{R} \mathbf{t} \|_p}{\|\mathbf{R} \mathbf{t^*} \|_p}
\end{equation}

where $\mathbf{t}^*$ is a vector representing the simple path durations calculated with all activities having their maximum duration. Having already established the usefulness of the nullspace, we can assert that if two different configurations, $\mathbf{t}_1$ and $\mathbf{t}_2$, differ by a vector $\boldsymbol{\delta}$ from the nullspace of $\mathbf{R}$, $N\left(\mathbf{R}\right)$, then they will exhibit the same stress. This follows trivially from Eq. \ref{eq:nullDelta}.

The special case $S_p = 1$ indicates that all simple paths have their maximum duration. The smaller the value of $S_p$, the greater the slack in the simple paths.
\subsection{Moore-Penrose inverse}
\subsubsection{Pseudoinverse}
The Moore-Penrose inverse, commonly known as the pseudoinverse, $\mathbf{R}^+$ of matrix $\mathbf{R}$, solves the least squares approximation of $\mathbf{R}\mathbf{t}=\boldsymbol{\tau}$:

\begin{equation}
\label{eq:tminsq}
\mathbf{t}_* = \mathbf{R}^+ \boldsymbol{\tau}
\end{equation}

This definition minimizes $\| \mathbf{R}\mathbf{t} - \boldsymbol{\tau} \|_2$ for a given $\boldsymbol{\tau}$, providing a particular vector $\mathbf{t}_*$ as a representative of those that best approximate $\boldsymbol{\tau} \approx  \mathbf{R}\mathbf{t}_* $. By Eq. \ref{eq:nullDelta}, any other vector differing from $\mathbf{t}_*$ by a vector $\boldsymbol{\delta}$ from  $N\left(\mathbf{R}\right)$ will equally provide the same $\boldsymbol{\tau}^*=\mathbf{Rt_*}$. 
\begin{example}
For the toy project, the pseudoinverse $\mathbf{R}^+$ results:
\[
\mathbf{R}^+ =\left[\begin{matrix} - \nicefrac{1}{8} & \nicefrac{1}{4} & \nicefrac{3}{8}\\\nicefrac{5}{8} & - \nicefrac{1}{4} & \nicefrac{1}{8}\\- \nicefrac{1}{4} & \nicefrac{1}{2} & - \nicefrac{1}{4}\\\nicefrac{1}{8} & - \nicefrac{1}{4} & \nicefrac{5}{8}\\\nicefrac{3}{8} & \nicefrac{1}{4} & - \nicefrac{1}{8}\end{matrix}\right]
\]
Given $\mathbf{t_1}=\begin{bmatrix} 5 & 5 & 2 & 5 & 5\\ \end{bmatrix}^T$ and $\mathbf{t_2}  = \begin{bmatrix} \frac{11}{2} &  \frac{9}{2} & 1 & \frac{9}{2} & \frac{11}{2}\\ \end{bmatrix}^T$ both differ by a vector in $N\left(\mathbf{R}\right)$, and thus both result in the same $\boldsymbol{\tau}=\begin{bmatrix} 10 & 12 & 10\\ \end{bmatrix}^T$. Applying the pseudoinverse to $\boldsymbol{\tau}$, Eq. \ref{eq:tminsq} gives a single vector as a result, which in this contrived example is $\mathbf{t}_2$.
 \end{example}

\subsubsection{Rank of $\mathbf{R}$}
Whether $\boldsymbol{\tau}^*$ is equal to $\boldsymbol{\tau}$ will depend on whether $\boldsymbol{\tau}$ belongs to the column space of $\mathbf{R}$. If the rank of $\mathbf{R}$ is equal to the number of simple paths, then any $\boldsymbol{\tau}$ will belong to the column space of $\mathbf{R}$ and will be reachable by an appropriated linear combination of the activities duration.  This fact is important so that the project manager may know in advance if there exist unreachable states.

\section{Related works}
\label{sec:mapping}
The ideas proposed in the current paper directly result from applying linear algebra techniques to Eq. \ref{eq:tauIsRt}. To assess whether this equation has ever been mentioned in the literature, we conducted the systematic mapping literature review described in the present section.

\subsection{Mapping}
Literature mapping techniques are useful at the beginning of a systematic literature review as a brainstorming and contextualization tool~\citep{cascade2012}. Mapping aims to summarize the state of the researched topic and to identify the volume of evidence, types of research applied, and/or available results in this area of research.

The methodology adopted is presented below, highlighting that in the very initial phase, a preliminary survey of bibliographic databases was conducted without following a systematic approach. This preliminary survey helped in detecting concepts and parameters used by other authors in this type of research work. Subsequently, a systematic search was carried out, allowing the team to extract conclusions that answered the research questions posed, ensuring that the research met all the requirements expected of this type of work.

It should be noted that the aim of the bibliographic search is not to turn this paper into a publication known as a systematic literature review on project duration determined as the product of the path matrix by the duration of activities. That might be a different publication, moving from the mapping conducted to the systematic review by applying quality parameters to the bibliographic findings on the topic.


The systematic mapping literature review methodology, also called literature mapping, is useful at the beginning of research for contextualizing ideas. Literature mapping aims to seek all the knowledge available about an idea and find the most relevant papers according to your research questions~\citep{JuanSLR, s22062160, fsedano20}.

As stated by \citet{fsedano20}: `The literature mapping process follows some steps which are practically the same as the SLR steps that will be described below. Before starting any literature mapping, the first step is to make a search on the Internet in order to verify if there already is a literature mapping about the intended topic'.

\subsubsection{Planning the Research Questions}
The definition of the research questions is the first step to take. The answers to these questions must fall within the field of the research being carried out~\citep{JuanSLR}.

Taking into account that the main context of this work is to determine whether the project duration has been calculated using matrix multiplication (path matrix times duration matrix) and if SVD has been employed in the matrix analysis for determining the relevance of paths and activities, the research questions are:
\begin {itemize}
\item RQ1: What methods exist for determining the duration of a project?
\item RQ2: What mathematical methods are used for project management? 
\end{itemize}

\subsubsection{Elaborating the \textsc{PICOC}}
The scope chosen for this work was as follows:
\begin {description}
\item[Population (\textsc{P}):] project duration 
\item[Intervention (\textsc{I}):] matrix pert cpm
\item[Comparison (\textsc{C}):] methods of determining project duration
\item[Outcome (\textsc{O}):] matrix product
\item[Context (\textsc{C}):] project management
\end{description}

\subsubsection{Inclusion and Exclusion Criteria}
Inclusion/exclusion criteria help define which articles are relevant to the study and can answer the research questions. For this systematic mapping (SMP), the inclusion criteria are:
\begin {description}
\item[IC1:] The paper proposes a method for calculating project duration;
\item[IC2:] The paper refers to matrices, even if they are of different types (incidence, DSM, MDM, mismatch, etc.);
\item[IC3:] The paper refers to a mathematical method for calculating duration;
\item[IC4:] The paper specifically talks about matrix of paths/routes;
\item[IC5:] The paper talks about the matrix of durations;
\item[IC6:] The paper discusses the importance of the different activities/pathways within the PERT diagram.
\end{description}
 The exclusion criteria are:
 \begin {description}
\item[EX0:] The work is not written in English;
\item[EX1:] Out of scope;
\item[EX2:] The paper talks about neural networks;
\item[EX3:] The paper focuses on EVM;
\item[EX4:] The paper refers to Markov chains;
\item[EX5:] There is an import error / The paper is not accessible;
\item[EX6:] The paper only discusses the PERT diagram and the concepts associated with the normal distribution.
\end{description}

\subsubsection{Source selection} 
 The subsequent decision involves selecting the databases from which to retrieve studies. This choice is critical as it can significantly influence the scope of the mapping exercise. For the purposes of this study, five of the most widely recognized and comprehensive libraries were considered, based on their relevance to the study's topics~\citep{fsedano20}:
 \begin {itemize}
\item SCOPUS (https://www.scopus.com/).
\item IEEE Digital Library (https://ieeexplore.ieee.org/).
\item Web of Science (www.webofknowledge.com).
\item Springer Link (http://link.springer.com).
\end{itemize}
The use of these sources is justified by their high degree of reliability and abundance of conferences and journals. These are significant archives.

\subsubsection{Creating the Search String and Choosing the Sources}
 The search string, also referred to as a query, is a construct that encapsulates all the key terms pertinent to the search. This string must be input into each selected database to identify relevant papers related to the theme. However, the exact format of the search string may vary across different databases, often requiring specific characters tailored to each platform.

The search string created for this work was (Base String):
( "duration" ) AND ( "project management" OR "project networks" ) AND ( "matrix" OR "matrix product" OR "DSM" OR "path" OR "cpm" OR "pert" ) AND NOT ( "construction" OR "fuzzy" OR "risk" OR "resource allocation" )

 \begin {itemize}
\item SCOPUS (https://www.scopus.com/). We have indicated that the search is only for the title, abstract and keywords of the paper: ( "duration" ) AND ( "project management" OR "project networks" ) AND ( "matrix" OR "matrix product" OR "DSM" OR "path" OR "cpm" OR "pert" ) AND NOT ( "construction" OR "fuzzy" OR "risk" OR "resource allocation" )
\item IEEE Digital Library. We have indicated that the search in “Abstract”, “Author Keywords” and “Title” in the same place as “All”: ( "duration" ) AND ( "project management" OR "project networks" ) AND ( "matrix" OR "matrix product" OR "DSM" OR "path" OR "cpm" OR "pert" ) AND NOT ( "construction" OR "fuzzy" OR "risk" OR "resource allocation" )
\item Web of Science (www.webofknowledge.com). The query was put in the search tab of the website and we added some terms: ( "duration" ) AND ( "project management" OR "project networks" ) AND ( "matrix" OR "matrix product" OR "DSM" OR "path" OR "cpm" OR "pert" ) NOT ( "construction" OR "fuzzy" OR "risk" OR "resource allocation" )
\item Springer Link (http://link.springer.com). The simple search on the website used exactly the same query quoted at the beginning: ( "duration" ) AND ( "project management" OR "project networks" ) AND ( "matrix" OR "matrix product" OR "DSM" OR "path" OR "cpm" OR "pert" ) NOT ( "construction" OR "fuzzy" OR "risk" OR "resource allocation" )
\end{itemize}

After the above mentioned search in the four databases, a second one is carried out, with the aim of contrasting the terms that appeared frequently in the previously selected articles and verifying the importance that other authors have given to the aspects that we are dealing with in this paper. 

For this final stage, which leads to new relevant papers to be studied, search strings are used:
\begin{itemize}
    \item SCOPUS: ( "duration" ) AND ( "project management" OR "project networks" ) AND ( "path" OR "cpm" OR "pert" ) AND ( "criticality" OR "cruciality" ) AND NOT ( "construction" OR "fuzzy" OR "risk" OR "resource allocation" )
    \item Web of Science: ("duration" ) AND ( "project networks" ) AND ("path" )AND (“criticality” OR “cruciality”) NOT ( "construction" OR "fuzzy" )
\end{itemize}

\subsubsection{Data Extraction}
Once we have all the articles of interest for the research, we extract the information from them in order to answer the research questions posed at the beginning of the mapping. 
The data extraction questions for this work are:
\begin {description}
\item[DE1:] How the paper calculates the project duration;
\item[DE2:] How the maximum project duration value is obtained;
\item[DE3:] How it defines the concepts of criticality and cruciality;
\item[DE4:] Which linear programming method is considered;
\item[DE5:] How it considers the activities and their duration;
\item[DE6:] What criterion is used to determine the influence of different activities on the critical path.
\end{description}

\tikzset{every node/.style={rounded corners,text badly centered,font=\small}}
\begin{figure}[!ht]
\caption{Mapping for theorical frame.}
\label{fig:mapping}
\centering
\resizebox{1\textwidth}{!}{%
\begin{tikzpicture}[>=latex]
	\tikzstyle{rect} = [draw, text width=3.5cm, minimum height=0.5cm, align=center, fill={rgb,255:red,235; green,235; blue,235}, text depth=0.5cm, text height=0.3cm]
	\tikzstyle{recty} = [draw,   rotate around={-270:(0,0)},  minimum width=2.5cm,  minimum height=1cm,  align=center, fill={rgb,255:red,255; green,249; blue,149}, text depth=0.5cm, text height=0.5cm, text width=2.5cm, inner sep=0 ]
	\tikzstyle{base} = [ draw, minimum width=8cm, minimum height=1cm, align=center,text depth=0.2cm, text height=0.2cm,  text width=10cm, fill=none ]
        
     \tikzstyle{base2} = [ draw, minimum width=5cm, minimum height=1.5cm, align=center, text depth=0.2cm, text height=-0.2cm,  text width=5cm, fill=none ]

    \node [rect] (scopus) at (2,13) {Scopus \\(n=213)};
    \node [rect] (ieee) at (6,13) {IEEE Digital Library\\(n=66)};
    \node [rect] (wos) at (10,13) {ISI Web of Science\\(n=89)};
    \node [rect] (springer) at (14,13) {Springer Link\\(n=55)};
    \node [rect] (prev) at (18,13) {Previous\\(n=74)};
    
    \node [recty] at (0,0) {Included};
    \node [recty] at (0,3) {Eligibility};
    \node [recty] at (0,6) {Screening};
    \node [recty] at (0,9) {Identification};
    
     \node [base] (excluded) at (10,9) {Record after duplicated removed \\ (n=408)};
     \draw[->] (scopus) -- (excluded);
     \draw[->] (ieee) -- (excluded);
     \draw[->] (wos) -- (excluded);
     \draw[->] (springer) -- (excluded);
     \draw[->] (prev) -- (excluded);

	
	\node [base] (criteria) at (10,6) {Records after apply selection criteria \\ (n=20)};
	\draw[->] (excluded) -- (criteria);
	
	\node [base] (eligibility) at (10,3) {Full-text articles assessed for eligibity \\ (n=54)};
	\draw[->] (criteria) -- (eligibility);
	
	\node [base] (quality) at (10,0) {Studies included after apply quality assessment criteria \\ (n=36)};
	\draw[->] (eligibility) -- (quality);

	\node [base2] (excluded2) at (17,7.5) {Records excluded \\ (n=388)};
	\draw[->] (10,7.5) -- (excluded2);
	\node [base2] (relevant) at (4.5,4.5) {Records derived from references in the relevant records\\(n=34)};
	\draw[->] (relevant) -- (10,4.5);
	\node [base2] (excluded3) at (17,1.5) {Full-text articles excluded, with reasons \\ (n=18)};
	\draw[->] (10,1.5) -- (excluded3);
\end{tikzpicture}
}
\end{figure}

As \citet{fsedano20} detailed, the process is the next one:
 \begin {itemize}
\item Identification: the papers found in each source using the query are saved and then the duplicate studies are removed.
\item Screening: just the title, abstract, and keywords are read applying the inclusion and exclusion criteria, the papers that are not approved by the criteria are removed too.
\item Eligibility: for the remaining articles, we applied the quality questions, so the papers need to be read fully in order to obtain the answers to those questions and a good score. The papers that do not have a score above the limit must be deleted.
\item Included: the papers with a high score are classified for the final review and we performed the data extraction using the data extraction form questions.
\end{itemize}

The tool used to perform this systematic mapping literature review was the Parsifal ~\citet{Parsifal}. It is good to organize the steps, plan the review, import the papers, answer the questions, and at the end generate a report about the review.

\subsection{Results}\label{sec:mapping_results}
The systematic mapping literature review process concluded that:
\begin{itemize}
    \item None of the papers dealt with the algebraic implications of Eq. \ref{eq:tauIsRt} presented in the current paper.
    \item Among the 408 papers scrutinized, only one \citep{CraigWSchmidt} referenced a linear transformation in a matrix akin to Eq. \ref{eq:tauIsRt}. The treatment of the transformation matrix, in that case, was limited to the square matrices case by augmenting the rectangular matrix using dummy paths and activities when needed.
\end{itemize} 

For the sake of completeness, we will mention here the final selection of papers, those whose aim was closer to the present paper, that resulted from the above-described process with a brief mention of the key ideas tackled in them.

The analysis of the results extracted from the mapping carried out will be grouped into four categories, determined according to the concepts addressed by the different papers. It should be noted that although many papers could be included in several of the categories created, they have only been included in one of them, as the aim of this section of the research work is to ascertain the state of the art of the subject.

The blocks that allow us to relate the conclusions derived from the analysis of the papers studied are as follows: Criteria 1, which includes those papers that study the stochastic and pert/cpm concepts. Criteria 2 includes the papers that analyse and highlight the path and path based concepts above others. Criteria 3 includes a brief analysis of the papers that attach importance to the concepts of criticality, cruciality and critical chain. Finally, Criteria 4 contains the conclusions of the study of a set of papers which, although they were selected because they responded to the research questions formulated in the mapping, could not be included in the three previous blocks because they emphasised other concepts, also linked to the subject matter, but not so focused on the concepts used to establish these categories. 

\subsubsection{Criteria 1: Stochastic Pert.}\label{subsec_criteria1}

The field of stochastic project evaluation and review technique (PERT) has undergone significant advancements, marked by the development of sophisticated methods to understand and model path criticality and project duration distributions under uncertainty.

As a first mention, it should be noted that  \cite{DODIN2006} delved into the path criticality index. By applying extreme value theory, Dodin achieved more precise estimates for the mean and variance of project durations, improving the reliability of project duration forecasts by accounting for extreme scenarios.

Building on this foundation, \cite{Klerides2010} introduced a two-stage decomposition-based stochastic programming model. This innovative approach, designed for project scheduling in environments with uncertain durations and time-cost settings, uses a path-based methodology to robustly manage these uncertainties, providing a valuable framework for project managers.

Meanwhile, \cite{Hayhurst1991} tackled the challenge of graph simplification to handle the exhaustive enumeration of paths. This breakthrough allowed for the exact solution of complex network problems, previously manageable only through approximation techniques, thereby enhancing the precision of project scheduling analyses.

The exploration continued with \cite{Monhor2011}, who investigated path criticality in stochastic PERT using a probabilistic approach. They defined the path criticality index for a path \(\pi_k\) as the probability \(p(\lambda(\pi_k) \geq \lambda(\pi_{k'})\)) for all simple paths \(\pi_{k'}\) different from \(\pi_k\). Recognizing the high computational cost of calculating this index due to the vast number of path durations requiring comparison, they introduced an approximation method via the probabilistic criticality index and the notion of the probabilistically critical path \citep{Monhor2011}.

In parallel, \cite{Li2014} explored the stochastic insuring critical path problem, advancing the understanding of risk management in project scheduling under uncertainty. This work complemented earlier efforts by providing insights into mitigating risks associated with critical path variability.

Further contributions came from \cite{Nasr}, who presented an efficient approach to accurately calculate the \(m\)-th moment of the duration of a stochastic network with feedback. This method is crucial for the precise modeling of project durations in dynamic environments, enhancing the ability to predict and manage project timelines.

Adding to these innovations, \cite{Chot1997} proposed the uncertainty-importance measure of activities (UIMA), also known as the 'cruciality' index. Evaluated using a modified Taguchi tolerance design technique, this measure helps in identifying and prioritizing activities that significantly impact the overall project duration under uncertainty.

Continuing the narrative, \cite{Vaseghi2024} studied the expected impact on the project duration distribution when corrective actions are taken on a selected set of activities. Their analytical risk analysis procedure aids in identifying critical activities whose timely correction can significantly mitigate project delays, offering a strategic tool for project managers.

\cite{Cho2004} contributed by defining a method for estimating the criticality index as the probability of an activity being on the critical path. This probabilistic measure is essential for understanding which activities are most likely to influence the project's completion time under uncertainty.

Finally, \cite{Pontrandolfo} examined the completion time of projects considering stochastic activity durations using PERT-state and PERT-path techniques. Their work provided valuable insights into managing and predicting project timelines in the presence of stochastic variables, completing the picture of advancements in this field.

Together, these studies form a comprehensive story of progress in stochastic PERT, equipping project managers with the tools and methodologies needed to navigate the uncertainties inherent in project scheduling and execution.

\subsubsection{Criteria 2: The path and path based.}\label{subsect_criteria2}

The study of critical path analysis has expanded significantly with the integration of fuzzy logic and stochastic methods, enhancing the precision and adaptability of project scheduling under uncertainty.

\cite{Chen2008} pioneered a fuzzy approach to critical path analysis, utilizing Yager's ranking method to order fuzzy sets. This innovative approach transformed the fuzzy critical path problem into a conventional one with crisp activity times, facilitating more straightforward application and analysis in project management.

In parallel, \cite{Copertari} developed a recursive technique for calculating project completion times in a polynomial number of steps, assuming beta distributions for all duration times. They introduced the concept of normalized criticality, defined as the normalized probability of an activity being the longest on a particular node. 

Complementing these efforts, \cite{Bruni2009} presented a method for estimating project completion times considering discrete random activity durations. Their work contributes to the body of knowledge by addressing the challenges of dealing with discrete stochastic variables in project scheduling.

\cite{Mummolo1997} further expanded the theoretical framework by defining uncertainty and criticality measures related to project evolutions in the context of the PERT-path network technique. These measures are essential for identifying and managing the impact of uncertainty on project timelines.

\cite{CraigWSchmidt} introduced a technique for computing the exact overall duration of a project when task durations have independent distributions. 

Meanwhile, \cite{Soroush1994} defined the 'most critical' path as the one with the lowest probability of completing its activities by a given time compared to every other path. 

These studies collectively advance the understanding and application of fuzzy and stochastic methods in critical path analysis. They provide project managers with sophisticated tools to handle the complexities and uncertainties inherent in project scheduling, ultimately leading to more robust and reliable project plans.

\subsubsection{Criteria 3: Cruciality / Criticality / Critical Chain}\label{subsection_criteria3}

The exploration of criticality in project management has evolved through various studies, each contributing unique perspectives and methodologies to better understand and manage critical activities and paths within projects.

\cite{Ammar2022} provided a comprehensive overview of different definitions and types of criticality for activities and paths, with a particular focus on fuzzy criticality measurement. This study helps in capturing the inherent uncertainties in project scheduling and offers a nuanced approach to criticality analysis.

\cite{Hajdu2016} advanced the field by studying the nine theoretical possible classes of critical activities. Their work contributes to a deeper understanding of the different ways activities can impact project schedules.

\cite{Bowers1996} explored the concepts of criticality and cruciality in both deterministic and stochastic contexts. 

In another significant contribution, \cite{Elmaghraby1999} investigated the sensitivity of project variability to the mean duration of activities. This study underscores the importance of understanding how changes in activity durations can affect the overall variability and reliability of project schedules.

\cite{Brozova2016} introduced a quantitative approach for identifying potentially dangerous tasks that could jeopardize the achievement of project objectives. 

Adding to the narrative, \cite{Soroush1994Heuristic} provided a heuristic approach for solving the optimal path problem through a deterministic network with a two-attribute fractional objective function. 

The concept of the critical chain, as discussed by \cite{Steyn}, further enriches the field by focusing on the integration of resource dependencies and project constraints. This approach emphasizes the importance of considering resource availability and project constraints in the determination of critical paths, thus providing a more realistic and comprehensive view of project scheduling.

Finally, \cite{Bianco2022} proposed a method to analyze criticalities and flexibilities within project schedules. 

These studies collectively enhance the theoretical and practical understanding of criticality in project management, providing project managers with advanced tools and methodologies to navigate and mitigate the complexities and uncertainties inherent in project scheduling.

\subsubsection{Criteria 4: Others.}\label{subsection_criteria4}
In addition to the primary criteria of stochastic PERT, fuzzy approaches, and criticality analysis, a diverse array of studies have significantly contributed to the broader thematic landscape of project management research. These works explore various aspects, including artificial intelligence, algorithm development, sensitivity analysis, and optimization techniques.
\cite{Hashfi2023} presented a comprehensive systematic literature review addressing the challenges and impacts of artificial intelligence on project management. Their review highlights the transformative potential of AI in enhancing project efficiency, decision-making, and overall management practices.

\cite{Qi2012} proposed an innovative algorithm for identifying hypo-critical paths based on the total float theorem. This algorithm provides a nuanced understanding of path criticality, offering project managers a tool to identify paths that, while not traditionally critical, can significantly affect project timelines under certain conditions.

From a sensitivity analysis perspective, \cite{Galvez2015} examined project uncertainty using a dependency structure matrix. Their study provides insights into how interdependencies between project activities can influence overall project risk and performance.

Focusing on construction schedules, \cite{Malyusz2021} analyzed the speed of different time analysis algorithms in networks characterized by maximal relationships. This work is crucial for improving the efficiency of scheduling algorithms in construction projects, which often involve complex and interdependent tasks.

In the manufacturing domain, \cite{Manzini2015} developed an approach to evaluate the criticality of missing components in a manufacturing-to-order production environment. This methodology helps in identifying and prioritizing critical components that can disrupt production schedules, thereby enhancing supply chain reliability.

\cite{Wood2020} integrated deterministic and stochastic calculations for project networks with parallel work pathways. This integration provides a comprehensive approach to project scheduling.

Investigating the forecasting accuracy of project durations, \cite{Elshaer2013} explored the impact of activity sensitivity measures using a dataset of 4,100 projects generated by the RanGen project network generator. Their findings underscore the importance of sensitivity measures in enhancing the reliability of project duration forecasts.

\cite{Bordley2019}, referencing \cite{demeulemeester2003rangen}, highlighted the importance of managing deadline uncertainty and numerically compared various crashing strategies. Their study offers practical insights into optimizing project schedules under tight deadlines.

\cite{Jevtic2011} utilized a modified shortest path algorithm within a project duration assessment model.

In the realm of resource-constrained project scheduling, \cite{Lombardi2012} proposed a min-flow algorithm for detecting Minimal Critical Sets using a Precedence Constraint Posting approach. Their use of the PSPlib dataset for numerical computations demonstrates the algorithm's efficacy in practical applications. Additionally, \cite{lombardi2009precedence} explored this approach with time lags and variable durations, further refining scheduling techniques under resource constraints.

Lastly, \cite{Sarjono2024} conducted a systematic literature review on optimizing project development implementation with project networks. Their review synthesizes various optimization strategies, providing a valuable resource for improving project execution efficiency.

These studies collectively enrich the field of project management by introducing innovative methodologies, algorithms, and insights that address a wide range of challenges and opportunities in project scheduling and execution.

\section{Discussion}
\label{sec:discussion}
\subsection{Criticality and cruciality}
The notion of which the most important paths and activities are depends on the complexity of the model we use to represent the actual behavior of the project. We will dissect this model into three different components.
\subsubsection{The topological component}
The first component just considers the graph topology of the project network as represented by $\mathbf{R}$. The SVD, as presented in this paper, is the tool to analyze the relative topological relevance of simple paths and activities. This helps identify which are the activities that appear in a larger fraction of simple paths, which paths condense the fundamental structure of the network, or the parts of the graph where bottlenecks might occur.

\subsubsection{The deterministic component}
By adding the second component of Eq. \ref{eq:tauIsRt}, the $\mathbf{t}$ vector of activity durations, we can analyze the resulting project duration as a projection of the vertices of the constraint equations, the route vectors, into a hyperplane whose slope is governed by the particular values in $\mathbf{t}$. The height of each vertex will vary depending on the particular set of $t_i$ values, thus providing a sense of relevance in terms of the duration of each simple path.

\subsubsection{The stochastic component}
Considering the stochastic nature of activity durations significantly impacts the relevance of project paths. This influence extends beyond mean values and standard deviations to encompass the project deadline itself~\citep{Soroush1994, Soroush1994Heuristic}. Consequently, the identification of dominant paths within a project configuration varies depending on the selected deadline, leading to distinct time intervals dominated by different simple paths.

Thus, the number of possible approaches for a project manager to act on the schedule of a project is diverse. 

\subsection{Simple paths, incidence matrix, and linear programming constraints}
Graph structure is usually condensed in incidence matrices whose columns represent the nodes and their rows represent the activities. Their contents are 0, 1, or -1 depending on whether the activity goes in, out, or does not involve a particular node. The incidence matrix ($\mathbf{H}$ in what follows) fully contains the information on the network structure.
Similarly, the constraint equations of the linear programming problem also fully represent the information on the network structure. It is not a coincidence that the transpose of the incidence matrix collects the coefficients on the left of the constraint equations.

The network routes, as encoded in $\mathbf{R}$, are the solutions to the constraint equations, and as such, they have been filtered by having to comply with the coefficients $b_i$ on the right of the constraint equations. As such, they are a discrete set of vectors belonging to the row space spanned by  $\mathbf{H}^T$, but there are other vectors in the row space of $\mathbf{H}^T$ that do not comply with the $b_i$ restriction.

Thus, we propose the SVD of  $\mathbf{R}$ and not of  $\mathbf{H}^T$ for two reasons. First, because the latter would provide information on the relevant activities and nodes, but not routes as they are not represented on either columns or rows of $\mathbf{H}$, and routes are much more meaningful than isolated nodes. Second, because the network structure represented on $\mathbf{R}$ complies with the conservation laws implied by the $b_i$ restrictions while $\mathbf{H}$ also accounts for unfeasible states.

Nevertheless, those interested in exploring the SVD of $\mathbf{H}$ should know that the rank of incidence matrices equals the number of nodes minus 1, leading to ill-conditioned matrices. Removing the column related to the first node of the project solves the problem by `grounding' that node, or taking it as a reference for the potential of the others.
\section*{Declarations of interest}
None.

\section*{Acknowledgements}
This research has been supported by grant TED2021-132356B-I00 funded by MCIN/AEI/10.13039/501100011033 and by the `European Union NextGeneration EU/PRTR', and the `European Union.-Next Generation UE/MICIU/Plan de Recuperacion, Transformacion y Resiliencia/Junta de Castilla y Leon.'
\section*{Contributions}
The authors contributed equally to all aspects of the research process, including idea generation, systematic mapping literature review process, draft writing, and review.

\bibliographystyle{chicago}
\bibliography{bibliography}

\begin{thebibliography}{}

\bibitem[\protect\citeauthoryear{Ammar and Abd-ElKhalek}{Ammar and
  Abd-ElKhalek}{2022}]{Ammar2022}
Ammar, M.~A. and S.~I. Abd-ElKhalek (2022).
\newblock Criticality measurement in fuzzy project scheduling.
\newblock {\em International Journal of Construction Management\/}~{\em 22},
  252--261.

\bibitem[\protect\citeauthoryear{Bajis}{Bajis}{2006}]{DODIN2006}
Bajis, D. (2006).
\newblock A practical and accurate alternative to pert.

\bibitem[\protect\citeauthoryear{Bianco, Caramia, and Giordani}{Bianco
  et~al.}{2022}]{Bianco2022}
Bianco, L., M.~Caramia, and S.~Giordani (2022, 4).
\newblock Project scheduling with generalized precedence relations: A new
  method to analyze criticalities and flexibilities.
\newblock {\em European Journal of Operational Research\/}~{\em 298}, 451--462.

\bibitem[\protect\citeauthoryear{Bordley, Keisler, and Logan}{Bordley
  et~al.}{2019}]{Bordley2019}
Bordley, R.~F., J.~M. Keisler, and T.~M. Logan (2019, 4).
\newblock Managing projects with uncertain deadlines.
\newblock {\em European Journal of Operational Research\/}~{\em 274}, 291--302.

\bibitem[\protect\citeauthoryear{Bowers}{Bowers}{1996}]{Bowers1996}
Bowers, J. (1996).
\newblock Identifying critical activities in stochastic resource constrained
  networks.

\bibitem[\protect\citeauthoryear{Brancali\~ao, Gon\c{c}alves, Conde, and
  Costa}{Brancali\~ao et~al.}{2022}]{s22062160}
Brancali\~ao, L., J.~Gon\c{c}alves, M.~A. Conde, and P.~Costa (2022).
\newblock Systematic mapping literature review of mobile robotics competitions.
\newblock {\em Sensors\/}~{\em 22\/}(6).

\bibitem[\protect\citeauthoryear{Brožová, Bartoška, Šubrt, and
  Rydval}{Brožová et~al.}{2016}]{Brozova2016}
Brožová, H., J.~Bartoška, T.~Šubrt, and J.~Rydval (2016).
\newblock Task criticalness potential: A multiple criteria approach to project
  management.
\newblock {\em Kybernetika\/}~{\em 52}, 558--574.

\bibitem[\protect\citeauthoryear{Bruni, Guerriero, and Pinto}{Bruni
  et~al.}{2009}]{Bruni2009}
Bruni, M.~E., F.~Guerriero, and E.~Pinto (2009, 10).
\newblock Evaluating project completion time in project networks with discrete
  random activity durations.
\newblock {\em Computers and Operations Research\/}~{\em 36}, 2716--2722.

\bibitem[\protect\citeauthoryear{Chen and Hsueh}{Chen and
  Hsueh}{2008}]{Chen2008}
Chen, S.~P. and Y.~J. Hsueh (2008, 7).
\newblock A simple approach to fuzzy critical path analysis in project
  networks.
\newblock {\em Applied Mathematical Modelling\/}~{\em 32}, 1289--1297.

\bibitem[\protect\citeauthoryear{Cho and Yum}{Cho and Yum}{2004}]{Cho2004}
Cho, J.~G. and B.~J. Yum (2004).
\newblock Functional estimation of activity criticality indices and sensitivity
  analysis of expected project completion time.
\newblock {\em Journal of the Operational Research Society\/}~{\em 55},
  850--859.

\bibitem[\protect\citeauthoryear{Chot and Yumj}{Chot and Yumj}{1997}]{Chot1997}
Chot, J.~G. and B.~J. Yumj (1997).
\newblock An uncertainty importance measure of activities in pert networks.

\bibitem[\protect\citeauthoryear{Conde-González, Rodríguez-Sedano,
  Fernández, Gonçalves, Lima, and García-Peñalvo}{Conde-González
  et~al.}{2020}]{fsedano20}
Conde-González, M., F.~Rodríguez-Sedano, C.~Fernández, J.~Gonçalves,
  J.~Lima, and F.~García-Peñalvo (2020, 10).
\newblock Fostering steam through challenge‐based learning, robotics, and
  physical devices: A systematic mapping literature review.
\newblock {\em Computer Applications in Engineering Education\/}~{\em 29}.

\bibitem[\protect\citeauthoryear{Copertari and Archer}{Copertari and
  Archer}{2001}]{Copertari}
Copertari, L. and N.~Archer (2001).
\newblock Calculating the theoretical project completion time of large networks
  in polynomial processing time.
\newblock In {\em PICMET '01. Portland International Conference on Management
  of Engineering and Technology. Proceedings Vol.1: Book of Summaries (IEEE
  Cat. No.01CH37199)}, Volume Supplement, pp.\  577--586 vol.2.

\bibitem[\protect\citeauthoryear{Cruz-Benito}{Cruz-Benito}{2016}]{JuanSLR}
Cruz-Benito, J. (2016, November).
\newblock Systematic literature review \& mapping.

\bibitem[\protect\citeauthoryear{Demeulemeester, Vanhoucke, and
  Herroelen}{Demeulemeester et~al.}{2003}]{demeulemeester2003rangen}
Demeulemeester, E., M.~Vanhoucke, and W.~Herroelen (2003).
\newblock Rangen: A random network generator for activity-on-the-node networks.
\newblock {\em Journal of scheduling\/}~{\em 6}, 17--38.

\bibitem[\protect\citeauthoryear{Elmaghraby, Fathi, and Taner}{Elmaghraby
  et~al.}{1999}]{Elmaghraby1999}
Elmaghraby, S.~E., Y.~Fathi, and M.~R. Taner (1999).
\newblock On the sensitivity of project variability to activity mean duration.
\newblock {\em Int. J. Production Economics\/}~{\em 62}, 219--232.

\bibitem[\protect\citeauthoryear{Elshaer}{Elshaer}{2013}]{Elshaer2013}
Elshaer, R. (2013).
\newblock Impact of sensitivity information on the prediction of project's
  duration using earned schedule method.
\newblock {\em International Journal of Project Management\/}~{\em 31},
  579--588.

\bibitem[\protect\citeauthoryear{Gálvez, Ordieres, and Capuz-Rizo}{Gálvez
  et~al.}{2015}]{Galvez2015}
Gálvez, R., R.~Ordieres, and R.~Capuz-Rizo (2015).
\newblock Analysis of project duration.

\bibitem[\protect\citeauthoryear{Hajdu, Skibniewski, Vanhoucke, Horvath, and
  Brilakis}{Hajdu et~al.}{2016}]{Hajdu2016}
Hajdu, M., M.~J. Skibniewski, M.~Vanhoucke, A.~Horvath, and I.~Brilakis (2016).
\newblock How many types of critical activities exist? a conjecture in need of
  proof.
\newblock Volume 164, pp.\  3--11. Elsevier Ltd.

\bibitem[\protect\citeauthoryear{Hashfi and Raharjo}{Hashfi and
  Raharjo}{2023}]{Hashfi2023}
Hashfi, M.~I. and T.~Raharjo (2023).
\newblock Exploring the challenges and impacts of artificial intelligence
  implementation in project management: A systematic literature review.
\newblock {\em International Journal of Advanced Computer Science and
  Applications\/}~{\em 14}, 366--376.

\bibitem[\protect\citeauthoryear{Hayhurst and Shier}{Hayhurst and
  Shier}{1991}]{Hayhurst1991}
Hayhurst, K.~J. and D.~R. Shier (1991).
\newblock A factoring approach for the stochastic shortest path problem.
\newblock {\em Operations Research Letters\/}~{\em 10}, 329--334.

\bibitem[\protect\citeauthoryear{Jevtic, Dobrilovic, Stojanov, and
  Stojanov}{Jevtic et~al.}{2011}]{Jevtic2011}
Jevtic, V., D.~Dobrilovic, J.~Stojanov, and Z.~Stojanov (2011).
\newblock Project duration assessment model based on modified shortest path
  algorithm and superposition.
\newblock pp.\  87--90. IEEE Computer Society.

\bibitem[\protect\citeauthoryear{Klerides and Hadjiconstantinou}{Klerides and
  Hadjiconstantinou}{2010}]{Klerides2010}
Klerides, E. and E.~Hadjiconstantinou (2010, 12).
\newblock A decomposition-based stochastic programming approach for the project
  scheduling problem under time/cost trade-off settings and uncertain
  durations.
\newblock {\em Computers and Operations Research\/}~{\em 37}, 2131--2140.

\bibitem[\protect\citeauthoryear{Li, Liu, and Yang}{Li et~al.}{2014}]{Li2014}
Li, Z., Y.~Liu, and G.~Yang (2014).
\newblock A new probability model for insuring critical path problem with
  heuristic algorithm.

\bibitem[\protect\citeauthoryear{Lombardi and Milano}{Lombardi and
  Milano}{2009}]{lombardi2009precedence}
Lombardi, M. and M.~Milano (2009).
\newblock A precedence constraint posting approach for the {RCPSP} with time
  lags and variable durations.
\newblock In {\em International Conference on Principles and Practice of
  Constraint Programming}, pp.\  569--583. Springer.

\bibitem[\protect\citeauthoryear{Lombardi and Milano}{Lombardi and
  Milano}{2012}]{Lombardi2012}
Lombardi, M. and M.~Milano (2012).
\newblock A min-flow algorithm for minimal critical set detection in resource
  constrained project scheduling.
\newblock {\em Artificial Intelligence\/}, 58--67.

\bibitem[\protect\citeauthoryear{Malyusz, Hajdu, and Vattai}{Malyusz
  et~al.}{2021}]{Malyusz2021}
Malyusz, L., M.~Hajdu, and Z.~Vattai (2021, 7).
\newblock Comparison of different algorithms for time analysis for cpm schedule
  networks.
\newblock {\em Automation in Construction\/}~{\em 127}.

\bibitem[\protect\citeauthoryear{Manzini and Urgo}{Manzini and
  Urgo}{2015}]{Manzini2015}
Manzini, M. and M.~Urgo (2015).
\newblock Critical components evaluation in manufacturing-to-order processes.
\newblock Volume~37, pp.\  146--151. Elsevier B.V.

\bibitem[\protect\citeauthoryear{Monhor}{Monhor}{2011}]{Monhor2011}
Monhor, D. (2011, 12).
\newblock A new probabilistic approach to the path criticality in stochastic
  pert.
\newblock {\em Central European Journal of Operations Research\/}~{\em 19},
  615--633.

\bibitem[\protect\citeauthoryear{Mummolo}{Mummolo}{1997}]{Mummolo1997}
Mummolo, G. (1997).
\newblock Measuring uncertainty and criticality in network planning by
  pert-path technique.

\bibitem[\protect\citeauthoryear{Nasr, Yassine, and Kasm}{Nasr et~al.}{}]{Nasr}
Nasr, W., A.~Yassine, and O.~A. Kasm.
\newblock An analytical approach to estimate the expected duration and variance
  for iterative product development projects.
\newblock {\em Research in Engineering Design\/}~{\em 27}.

\bibitem[\protect\citeauthoryear{Parsifal}{Parsifal}{}]{Parsifal}
Parsifal.
\newblock Parsifal.

\bibitem[\protect\citeauthoryear{Peñalvo}{Peñalvo}{2017}]{cascade2012}
Peñalvo, D. F. J.~G. (2017, 1).
\newblock Revisión sistemática de literatura para artículos.

\bibitem[\protect\citeauthoryear{Pontrandolfo}{Pontrandolfo}{}]{Pontrandolfo}
Pontrandolfo, P.
\newblock Project duration in stochastic networks by the pert-path technique.

\bibitem[\protect\citeauthoryear{Qi and Zhao}{Qi and Zhao}{2012}]{Qi2012}
Qi, J. and X.~Zhao (2012).
\newblock Algorithm of finding hypo-critical path in network planning.
\newblock {\em Physics Procedia\/}~{\em 24}, 1520--1529.

\bibitem[\protect\citeauthoryear{Sarjono and Kurnia}{Sarjono and
  Kurnia}{2024}]{Sarjono2024}
Sarjono, H. and V.~D. Kurnia (2024).
\newblock A systematic literature review: Optimization of implementation of
  project development in the company with pert and cpm method.

\bibitem[\protect\citeauthoryear{Schmidt and Grossmann}{Schmidt and
  Grossmann}{1998}]{CraigWSchmidt}
Schmidt, C.~W. and I.~E. Grossmann (1998).
\newblock The exact overall time distribution of a project with uncertain task
  durations.

\bibitem[\protect\citeauthoryear{Soroush}{Soroush}{1994a}]{Soroush1994}
Soroush, H.~M. (1994a).
\newblock The most critical path in a pert network.

\bibitem[\protect\citeauthoryear{Soroush}{Soroush}{1994b}]{Soroush1994Heuristic}
Soroush, H.~M. (1994b).
\newblock The most critical path in a pert network: A heuristic approach.

\bibitem[\protect\citeauthoryear{Steyn}{Steyn}{}]{Steyn}
Steyn, H.
\newblock An investigation into the fundamentals of critical chain project
  scheduling.

\bibitem[\protect\citeauthoryear{Vaseghi, Martens, and Vanhoucke}{Vaseghi
  et~al.}{2024}]{Vaseghi2024}
Vaseghi, F., A.~Martens, and M.~Vanhoucke (2024).
\newblock Analysis of the impact of corrective actions for stochastic project
  networks.
\newblock {\em European Journal of Operational Research\/}~{\em 316}, 503--518.

\bibitem[\protect\citeauthoryear{Wood}{Wood}{2020}]{Wood2020}
Wood, D.~A. (2020).
\newblock High-level stochastic project cost and duration planning methodology
  integrating earned duration, schedule and value, criticality, cruciality and
  downside risk metrics.
\newblock {\em International Journal of Operational Research\/}~{\em 39},
  160--204.

\end{thebibliography}
\end{document}